\documentstyle[epsf, rotate]{l-aa}
\begin{document}

\thesaurus{11.01.2, 11.17.4, 13.07.2, 13.07.3}
\title{Spectral Variability in PKS 0528+134 at Gamma-Ray Energies}
\author{M.  B\"ottcher \inst{1} \and W. Collmar \inst{2}}
\institute{Rice University, Space Physics and Astronomy Department,
6100 S. Main Street, Houston, TX 77005 -- 1892, USA
\and
Max-Planck-Institut f\"ur Extraterrestrische Physik, Postfach 16 03,
D --- 85 740 Garching, Germany }
\date{Received 25 September 1997; accepted 14 October 1997}
\offprints{M. B\"ottcher}

\maketitle
\markboth{M. B\"ottcher and W. Collmar: Spectral Variability in PKS 
0528+134 at Gamma-Ray Energies}{M. B\"ottcher and W. Collmar: 
Spectral Variability in PKS 0528+134 at Gamma-Ray Energies}

\begin{abstract}
We present model calculations to the observed spectral variability
in the $\gamma$-ray spectrum of the ultraluminous blazar-type quasar
PKS~0528+134. We argue that the observed appearance of a spectral 
break between the COMPTEL and the EGRET energy range can plausibly
be explained by a variation of the Doppler beaming factor in the
framework of a relativistic jet model for AGNs, combining the ERC 
and the SSC model. This may be an instructive example for the
different dependence of these two spectral components on the
Doppler factor, as predicted in earlier work.
\end{abstract}

\section{Introduction}

Up to now, more than 50 blazar-type AGN have been detected by EGRET 
as emiters of high-energy $\gamma$-rays above 100 MeV (Mattox et al. 
1997). These sources are identified with flat-spectrum ($\alpha$ $>$ 
-0.5) radio sources classified as QSOs and BL Lac objects.
A large fraction of these blazars exhibits variability at
$\gamma$-ray energies on time scales of days to months
(Mukherjee et al. 1997).
At radio wavelengths, all blazars 
can be recognized as bright, compact sources with a flat 
synchrotron spectrum emanating from outflowing plasma jets 
that are nearly aligned with our line of sight. Relativistic
beaming is required in the objects in view of the luminosity 
and variability time scales (Dermer and Gehrels 1995), in 
accord with VLBI observations indicating that superluminal 
motion is a common feature in this class of AGN 
(e. g., Pohl et al. 1995, Barthel et al. 1995, 
Krichbaum et al. 1995).

One of the most luminous examples of this class of objects
is the quasar PKS~0528+134. It is known as a bright radio 
source with a flat radio spectrum. Recent VLBI images show 
a one-sided core-jet structure of $\sim$~5 mas length and 
superluminal motion (Pohl et al. 1995). An optical spectrum, 
obtained in November 1991 with the ESO 3.6~m telescope at 
La Silla / Chile, revealed a redshift of z = 2.07 (Hunter 
et al. 1993). At X-rays, the source was first detected with 
the Einstein observatory (Bregman et al. 1985). At $\gamma$-ray 
energies, PKS~0528+134 is detected by EGRET, COMPTEL and OSSE 
aboard the Compton Gamma-ray Observatory (CGRO) (Mukherjee et al. 
1996, Collmar et al. 1997, McNaron-Brown et al. 1995), showing 
a time-variable $\gamma$-ray flux with intensity variations up 
to a factor of $\sim$20 (Mukherjee et al. 1996).
The broadband spectrum of PKS~0528+134 reveals that during 
$\gamma$-ray high states its total energy output is dominated 
by its $\gamma$-ray emission. Under the assumption 
of isotropic emission, the total luminosity of the object
during $\gamma$-ray high states would be $L_{iso} \approx 
10^{49}$~ergs~s$^{-1}$ (throughout this paper, we use 
$H_0 = 75$~km~s$^{-1}$~Mpc$^{-1}$ and $q_0 = 0.5$).

The strongest EGRET blazar detections can be characterized
by a single power-law spectrum with differential photon
spectral indices $1.5 \le s \le 2.7$ (Thompson et al.
1995). In some cases, such as PKS~0528+134 (Mukherjee
et al. 1996), the EGRET spectrum (30~MeV - $\sim$10~GeV) 
seems to harden during $\gamma$-ray flares. A recent analysis 
of the first 3.5 years of COMPTEL observations of PKS~0528
+134 (Collmar et al. 1997) indicated a correlation between its 
$\gamma$-ray flux and the occurence of a spectral break at 
MeV energies. During states of high $\gamma$-ray intensity 
the combined EGRET and COMPTEL spectra
($\sim$1~MeV to $\sim$10~GeV) are characterized by 
a spectral break between the EGRET and the COMPTEL 
energy range, while during $\gamma$-ray low states
the $\gamma$-ray spectrum from MeV to GeV energies 
is consistent with a single power-law.

Measurements of broadband spectral variability in 
$\gamma$-ray blazars can put the most severe constraints
on models for these objects. By now, it is widely
accepted that the high-energy emission of flat-spectrum radio quasars
and BL Lac objects is produced by inverse-Compton
scattering of soft radiation off ultrarelativistic
electrons (and positrons) in a relativistic jet
directed at a small angle with respect to our line 
of sight (e. g. Schlickeiser 1996). A major subject
of debate is the source of the soft photons which are
upscattered to $\gamma$-ray energies. These can be
synchrotron photons produced by the jet itself
(the SSC model, e. g. Maraschi et al., 1992, 
Bloom \& Marscher 1996), accretion disk 
radiation entering the jet directly (the ERC
model, Dermer, Schlickeiser \& Mastichiadis 1992,
Dermer \& Schlickeiser 1993) or after being
rescattered by the broad line regions (Sikora, 
Begelman \& Rees 1994), or jet radiation being 
reflected by a torus of cold material and 
reentering the jet (Ghisellini \& Madau 1996).

Detailed investigations of the ERC model and
of the SSC model have recently been presented
by Dermer, Stur\-ner \& Schlickeiser (1997) and
by Bloom \& Marscher (1996), respectively. These
reveal that sharp spectral breaks in the $\gamma$-ray
spectra of blazars at several MeV favor the ERC
mechanism as a possible explanation because such 
breaks cannot be produced by a homogeneous SSC 
model. In contrast, a smooth transition from
a hard spectrum at X-ray and soft $\gamma$-ray 
energies to a softer EGRET spectrum seems indicative 
of SSC being the dominant radiation mechanism.
Nevertheless, a pure version of one of these
models seems to be an unrealistic simplification
since both of these radiation mechanisms are
plausibly at work in relativistic AGN jets.
A synthesis of both models has been investigated
in detail by B\"ottcher, Mause \& Schlickeiser 
(1997).

Dermer et al. (1997) have shown that the 
ERC radiation exhibits a stronger dependence
($F_{ERC} (\epsilon) \propto D^{3 + s}$)
on the Doppler factor $D = (\Gamma \, [1 - \beta_{\Gamma}
\cos\theta_{obs}])^{-1}$ than the SSC component ($F_{SSC}
(\epsilon) \propto D^{(5 + s)/2}$). Here, $\Gamma
= (1 - \beta_{\Gamma})^{-1/2}$ is the bulk Lorentz
factor of the ultrarelativistic jet plasma and $s$ is 
the spectral index of the initial particle distribution 
function. The kind of spectral variability as observed in 
PKS~0528+134 can therefore easily be explained
by a change in the Doppler factor (i. e. a change
in the bulk Lorentz factor or in the angle $\theta_{obs}$
between the direction of motion and the line of sight) 
in the framework of a combined SSC/ERC model. 
According to this idea, we suggest that in the case
of PKS~0528+134 the SSC radiation dominates 
the $\gamma$-ray spectrum in the $\gamma$-ray low 
state (with no spectral break in the COMPTEL regime 
required), while in the $\gamma$-ray high state
an additional ERC component becomes dominant in
the $\gamma$-ray spectrum at energies around 
10 MeV. In this Letter, we present model calculations
demonstrating the viability of this idea.

\section{The model}

To produce spectral fits to PKS~0528+134 we use the combined
ERC/SSC model described in detail by B\"ottcher et al. (1997). 
We assume a spherical blob filled with ultrarelativistic pair 
plasma which is moving out along an existing jet structure 
perpendicular to an accretion disk around a black hole of 
$M = 5 \cdot 10^{10} M_{\odot}$. The luminosity of the 
accretion disk is assumed to be $L_0 = 5 \cdot 10^{46} 
{\rm erg \, s}^{-1}$. The blob moves with a constant bulk 
Lorentz factor $\Gamma$. The particles inside the blob 
are isotropically distributed according to a power-law 
$n(\gamma) \propto \gamma^{-s}$ for $\gamma_1 \le \gamma 
\le \gamma_2$ in the rest frame of the blob. We assume 
reacceleration in the initial phase to be inefficient 
in comparison to radiative cooling of the pairs and follow 
the self-consistent evolution of the pair distribution 
as the blob moves out, taking into account synchrotron 
radiation, SSC scattering to arbitrarily high order, ERC 
scattering, $\gamma$-$\gamma$ absorption and pair production.
Comparison to observations is made on the basis of the
time-averaged flux of a blob since the inferred cooling
timescale of the material in the blob (in the observer's
frame) is shorter than the time resolution of COMPTEL
and EGRET.

In our model we assume the ejection of individual
plas\-ma blobs with different bulk Lorentz 
factors and slight differences in the internal energy
distributions of the material in the blob. These are described
by the following parameters: $\gamma_{1,2}$ (the low- and 
high-energy cutoffs in the Lorentz factors of the electrons), 
s (the power-law index of the electon distribution), $B$ 
(the magnetic field), $\Gamma$ (the bulk Lorentz factor
of the blob), $\theta_{obs}$ (the angle between the line of
sight and the jet axis), $z$ (the starting height of the
center of the blob above the accretion disk).

\begin{figure}
\rotate[r] {
\epsfxsize=6cm
\epsffile[100 20 600 50] {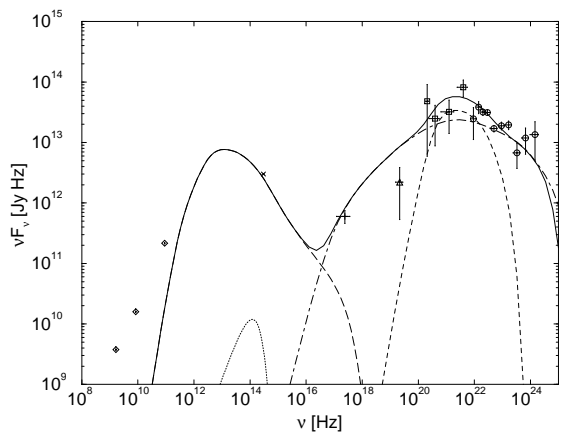} }
\caption[]{Fit to the (non simultaneous) broadband spectrum of
PKS~0528+134. Parameters: $\gamma_1 = 2 \cdot 10^3$, $\gamma_2 
= 9 \cdot 10^4$, $s = 3.0$, $n = 60 \, {\rm cm}^{-3}$, $R_B = 
4 \cdot 10^{16} \, {\rm cm}$, $B = 1.3$~G, $z = 4 \cdot 10^{-1}$~pc, 
$L_0 = 5 \cdot 10^{46} \, {\rm erg \, s}^{-1}$, $\Gamma = 20$, 
$\theta = 2.5^{\circ}$. Dashed: ERC, dot-dashed: SSC, 
long dashed: synchrotron, dotted: accretion disk, solid: total}
\end{figure}

\section{Spectral fits}

In this section, we compare the results of model 
calculations to the spectra of PKS~0528+134 demonstrating 
the viability of the combined ERC/SSC model to explain
the broadband spectrum as well as the peculiar spectral
variability of this object. As a first step, we produce a fit
to the (non simultaneous) broadband spectrum of PKS~0528+134
which is shown in Fig. 1. The radio data are taken from
Reich et al. (1993). The optical data point is still under
debate because PKS~0528+134 is located behind a strongly
absorbing cloud of interstellar material, and the correct
dereddening factor is unclear (Wagner 1997, private communication).
Here, we use the same optical 
point as used by Mukherjee et al. (1996). The ROSAT point
is taken from Zhang et al. (1994), and the OSSE data point 
is from McNaron-Brown et al. (1995). The contemporaneous
COMPTEL and EGRET data from Viewing Period (VP)~0 (April 1991) 
are taken from Collmar et al. (1997). 

Even though the broadband spectrum shown in Fig. 1 is not
measured simultaneously, it gives us an idea of the parameters 
to be used in the detailed fits to the $\gamma$-ray spectra
in the high and low states of PKS~0528+134, as presented by
Collmar et al. (1997). In the model described above, it
is assumed that the radio spectrum (below $\sim 100$~GHz)
is predominantly produced in the more distant parts of 
the jet, because the $\gamma$-ray emitting plasma is 
optically thick to synchrotron self-absorption at radio 
frequencies. The OSSE data point is not reproduced by 
our fit because it has not been measured contemporaneously
with the $\gamma$-ray spectrum and does therefore 
not correspond to the $\gamma$-ray high state.
(Figure 2, however, illustrates that our spectral fit to
the $\gamma$-ray low state is in good agreement
with the OSSE spectrum at that time.)

Fig. 1 demonstrates that the spectral break between 
the COMPTEL and the EGRET energy range can well be 
explained by an ERC component dominating the spectrum 
at energies around 10 MeV in the $\gamma$-ray high state. 
The strong spectral break being located at $\sim 10$~MeV
constrains the input soft photons to the component
causing this break to a relatively low energy. Given
the required total luminosity of the disk photon field,
this leads us to the conclusion of an accretion disk 
radiates at only $\approx 1$~\% of the Eddington luminosity 
for a central black-hole mass of $\sim 5 \cdot 10^{10} \, M_{\odot}$,
which is consistent with dynamical studies of the central
region of PKS~0528+134. The size of the emitting region
of $R_B = 4 \cdot 10^{16}$~cm is consistent with the 
timescale of $\sim 2$~days for intensity variations of 
order 100 \% as found recently by Wagner et al. (1997).

\begin{figure}
\rotate[r] {
\epsfxsize=6cm
\epsffile[100 20 600 50] {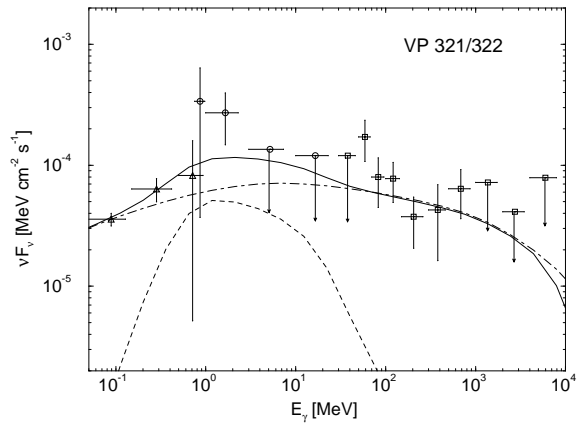} }
\caption[]{Fit to the $\gamma$-ray (OSSE: triangles, 
COMPTEL: circles, EGRET: squares) spectrum of PKS~0528+134 
in its low state (combined VP 321/322). Parameters
are the same as for Fig. 1, except $\Gamma = 7$, $s = 2.8$. 
Dashed: ERC, dot-dashed: SSC, solid: total}
\end{figure}

The low state, as shown in Fig. 2, could well be reproduced by
a pure SSC model. For our fit to this state we used basically 
the same set of parameters as for the high state (Fig. 1), but 
a lower bulk Lorentz factor ($\Gamma = 7$ in the low state 
versus $\Gamma = 20$ in the high state). The different
dependence of the SSC and the ERC component on the Doppler
factor (Dermer et al. 1997) then causes the observed 
spectral variability in the $\gamma$-ray regime. Since 
we have no simultaneous observations at other wavelength
bands, the fit to the non simultaneous radio to soft X-ray
spectrum using our parameters for the low state (which
is equally good as the one shown in Fig. 1) does not
yield any physically relevant information. In all 
figures the SSC and ERC components are plotted without 
taking $\gamma$-$\gamma$ absorption into account. In 
the calculation of the total spectrum as well as for 
the simulation of the evolution of the jet material, 
$\gamma$-$\gamma$ absorption and pair production 
are included self-consistently. This is the reason
for the (unabsorbed) SSC contribution being higher
than the total escaping flux at high energies in
the figures.

\begin{figure}
\rotate[r] {
\epsfxsize=6cm
\epsffile[100 20 600 50] {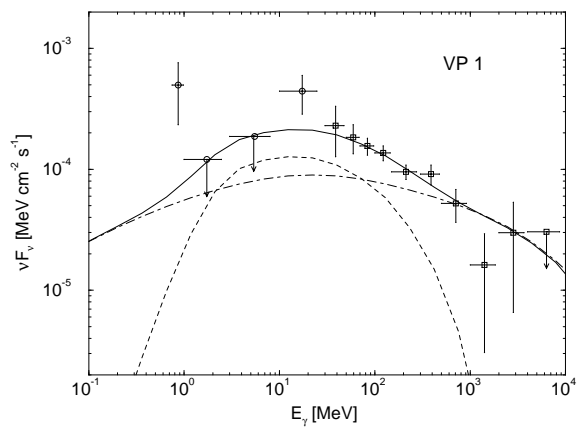} }
\caption[]{Fit to the $\gamma$-ray (COMPTEL and EGRET) spectrum 
of PKS~0528+134 in its high state in VP 1. Parameters are
the same as for Fig. 1, except $\gamma_1 = 2.4 \cdot 10^3$, 
$\gamma_2 = 8 \cdot 10^4$, $s = 3.1$, $\Gamma = 22$. 
Dashed: ERC, dot-dashed: SSC, solid: total}
\end{figure}

Figs. 3 and 4 illustrate that slight changes in the intrinsic 
particle distribution functions and the bulk Lorentz factor 
as compared to the fit shown in Fig. 1, lead to a good fit
to the $\gamma$-ray high states observed in viewing period
1, while the state showing the highest flux in the history
of these observations, found in VP~213, seems indeed to be
exceptional and requires a very high bulk Lorentz factor
and a significantly harder particle spectrum than the other
states. Its $\gamma$-ray spectrum could equally well be
reproduced with a pure ERC model, thus confirming the
trend that the more luminous the object is in $\gamma$-rays
the more dominant is the ERC component. The relevant parameters 
of the different fits are given in Table 1. 

\begin{figure}
\rotate[r] {
\epsfxsize=6cm
\epsffile[100 20 600 50] {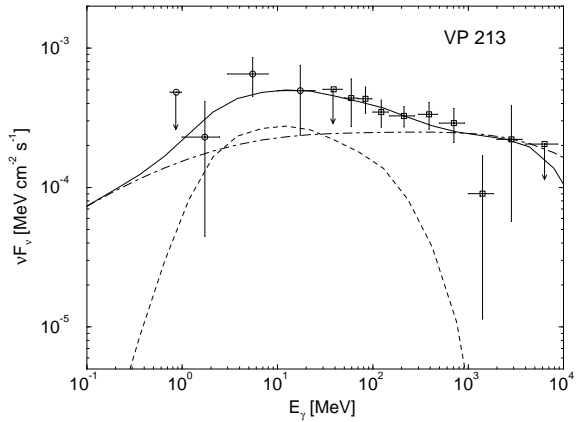} }
\caption[]{Fit to the $\gamma$-ray (COMPTEL and EGRET) spectrum 
of PKS~0528+134 in its high state in VP 213. Parameters are
the same as for Fig. 1, except $\gamma_2 = 10^5$, $s = 2.4$, 
$\Gamma = 25$. Dashed: ERC, dot-dashed: SSC, solid: total}
\end{figure}

\begin{table}[hbtp]
\caption[]{Interesting model parameters for the different observational
periods. $F_{1 - 10^4}$ is the integrated flux $F_{\nu}$ from 1~MeV
to 10~GeV of our model fits in units of keV~cm$^{-2}$~s$^{-1}$. The
different $\gamma$-ray fluxes reflect mainly the different Lorentz
factors $\Gamma$ of the individual blobs.}

\begin{center}\begin{tabular}{ccccc}
\hline
 Parameter   & VP 321/322     & VP 0           &  VP 1   & VP 213   \\
\hline
 $\gamma_1$  & $2 \cdot 10^3$ & $2 \cdot 10^3$ & $2.4 \cdot 10^3$ & $2 \cdot 10^3$ \\
 $\gamma_2$  & $9 \cdot 10^4$ & $9 \cdot 10^4$ & $8 \cdot 10^4$   & $10^5$         \\
  s          & 2.8            & 3.0            &  3.1             & 2.4            \\
 $\Gamma$    & 7              & 20             &  22              & 25             \\
\hline
 $F_{1 - 10^4}$ & 0.6         & 1.1            & 1.1              & 3.3            \\
\hline
\end{tabular}\end{center}\end{table}

\section{Summary and conclusions}

We presented model calculations on the spectral variability
observed in the MeV to GeV spectrum of the ultraluminous
blazar-type quasar PKS~0528+134. This quasar exhibits a 
peculiar spectral behavior between its $\gamma$-ray low
and high states, indicative of an additional spectral
component showing up around several MeV in the high state. 

Our model fits demonstrate that a combined ERC / SSC model
is very well suited to reproduce this behavior, based on the
different dependence of the ERC and the SSC component on the
Doppler boosting factor. In the $\gamma$-ray low state, the
$\gamma$-ray spectrum can basically be reproduced as a pure
SSC spectrum, while in the $\gamma$-ray high state, due
to an increase in the bulk Lorentz factor of the jet, the
ERC component becomes dominant over a significant part
of the $\gamma$-ray spectrum, producing a spectral break
around several MeV.

The spectral variability in PKS~0528+134 appears to be an 
interesting manifestation of the different Doppler boosting 
patterns of the ERC and the SSC radiation, as predicted
earlier. A similar variation of the Doppler factor can,
of course, also result from a change in the viewing angle.
For example, the reduction of the beaming factor as appropriate
to explain the spectrum of VP~321/322 (Fig. 2) can equally
result from assuming a viewing angle of 4.7$^{\circ}$ (instead
of 2.5$^{\circ}$ as used for our simulations) when leaving the
bulk Lorentz factor constant at $\Gamma = 20$. The diagnostic
on the Doppler factor presented in this Letter cannot distinguish 
between these two effects.


\begin{thebibliography}{}

\bibitem[]{} Barthel P.D. et al., 1995, ApJ 444, L21

\bibitem[]{} Bloom, S. D., Marscher, A. P., 1996, ApJ 461, 657

\bibitem[]{} B\"ottcher, M., Mause, H., \& Schlickeiser, R., 1997,
A \& A, 324, 395

\bibitem[]{} Bregman, J. N., Glassgold, A. E., Huggins, P. J., 
et al., 1985, ApJ 291, 505

\bibitem[]{} Collmar, W., Bennett, K., Bloemen, H., et al., 
1997, A\&A, in press

\bibitem[]{} Dermer, C. D., Gehrels N., 1995, ApJ 447, 103

\bibitem[]{} Dermer, C. D., Schlickeiser, R., Mastichiadis, A., 1992,
A\&A 256, L27

\bibitem[]{} Dermer, C. D., Schlickeiser, R., 1993, ApJ 416, 458

\bibitem[]{} Dermer, C. D., Sturner, S. J., Schlickeiser, R., 1997
ApJS 109, 103

\bibitem[]{} Ghisellini, G., Madau, P., 1996, MNRAS 280, 67

\bibitem[]{} Hartman, R. C., Webb, J. R., Marscher, A. P.,
et al., 1996, ApJ 461, 698

\bibitem[]{} Hunter, S. D., Bertsch, D. L., Dingus, B. L.,
et al., 1993, ApJ 409, 134

\bibitem[]{} Krichbaum T.P. et al., 1995, in {\it Quasars and AGN}, Eds. 
M. Cohen \& K. Kellermann, Proc. Nat. Acad. Sci. 92, 11377

\bibitem[]{} Maraschi, L., Ghisellini, G., Celotti, A., 1992, ApJ 397, L5

\bibitem[]{} Mattox, J. R., Schachter, J., Molnar, L., et al.,
1997, ApJ 481, 95

\bibitem[]{} McNaron-Brown K., Johnson, W. N., Jung, G. V., 
et al., 1995, ApJ 451, 575

\bibitem[]{} Mukherjee, R., Dingus, B. L., Gear, W. K., et al.,
1996, ApJ 470, 831

\bibitem[]{} Mukherjee R. et al., 1997, ApJ submitted

\bibitem[]{} Pohl M., Reich, W., Krichbaum, T., et al., 1995, 
A\&A 303, 383

\bibitem[]{} Reich, W., Steppe, H., Schlickeiser, R., et al.,
1993, A\&A 273, 65

\bibitem[]{} Schlickeiser, R., 1996, Space Sci. Rev. 75, 299

\bibitem[]{} Sikora, M., Begelmann, M. C. \& Rees, M. J.,
1994, ApJ 421, 153

\bibitem[]{} Thompson D.J., Bertsch, D. L., Dingus, B. L.,
et al., 1995, ApJS 102, 259

\bibitem[]{} Wagner, S. J., v. Montigny, C., \& Herter, M., 1997,
4th Compton Symposium, in press

\bibitem[]{} Zhang, Y. F., et al., 1994, ApJ 432, 91

\end{thebibliography}
\end{document}